\documentclass[preprint,aps]{revtex4}
\newcommand{\be}{\begin{equation}}
\newcommand{\ee}{\end{equation}}
\newcommand{\ba}{\begin{eqnarray}}
\newcommand{\ea}{\end{eqnarray}}

\begin{document}

\draft

\title{Brans-Dicke Theory as a Unified Model \\
       for Dark Matter - Dark Energy} 
       
\author{Hongsu Kim\footnote{e-mail : hongsu@astro.snu.ac.kr}}

\address{Astronomy Program, SEES, Seoul National University, Seoul, 151-742, KOREA and \\
International Center for Astrophysics \\
Korea Astronomy and Space Science Institute, Daejeon, 305-348, KOREA\footnote{present address}}


\begin{abstract}
The Brans-Dicke (BD) theory of gravity is taken as a possible theory of k-essence coupled to gravity. 
It then has been realized that the BD scalar field does indeed play a role of a k-essence, 
but in a very unique way which distinguishes it from other k-essence fields studied in the literature. 
That is, first in the BD scalar field-dominated era when the contribution from this k-essence overwhelms
those from other types of matter, the BD theory predicts the emergence of a yet-unknown 
{\it zero acceleration} epoch which is an intermediate stage acting as a ``crossing bridge'' between the 
decelerating matter-dominated era and the accelerating phase. Upon realizing this, next, closer
study of the effects of this k-essence on the evolutionary behavior of the matter-dominated and the
accelerating eras has been performed. The result of the study indicates that the BD scalar field
appears to interpolate {\it smoothly} between these two late-time stages by speeding up the expansion
rate of the matter-dominated era somewhat while slowing down that of the accelerating phase to some
degree. Thus with the newly found BD scalar field-dominated era in between these two, the late-time
of the universe evolution appears to be mixed sequence of the three stages.

\end{abstract}


\maketitle

\narrowtext

\newpage
\begin{center}
{\rm\bf I. Introduction}
\end{center}

Perhaps one of the greatest challenges in the theoretical cosmology today would be to understand
the emergence and nature of the observed late-time acceleration of the universe and provide an answer
to the so-called ``cosmic coincidence conundrum'' which concerns the puzzle :
why at the present epoch are the energy densities of dark energy and of dust-like dark matter
of the same order of magnitude ? For instance, according to the recent WMAP data \cite{wmap}, the universe
energy density appears to consist of approximately 4-per cent of that of visible matter, 21-per cent of that of
dark matter and 75-per cent of that of dark energy. Up until now, the most conservative candidate for the
dark energy is the cosmological constant $\Lambda$ and perhaps the most fashionable candidates with
non-trivial dynamics are quintessence and k-essence. And the main difference
between the two models is that the standard quintessence models \cite{quintessence} involve canonical kinetic terms
and the sound speed of $c^2_{s} = 1$ while the k-essence models \cite{kessence} employ rather exotic scalar fields
with non-canonical (non-linear) kinetic terms which typically lead to the {\it negative} pressure.
And the most remarkable property of these k-essence models is that the typical k-essence field can
overtake the matter energy density and induce cosmic acceleration only at the onset of the 
matter-dominated era for some period, at about the present epoch. Usually the emergence of such exotic 
type of scalar fields with non-canonical kinetic terms has been attributed to the string/supergravity
theories in which the non-linear kinetic terms generically appear in the effective action describing
moduli and massless degrees of freedom. We also note that a pioneering work on accelerating
cosmology which has some relevance to the current intensive study of scalar field models for dark energy
had been presented by Peebles and Ratra \cite{peebles1} some time ago. 
In this regard, here we focus particularly on the Brans-Dicke
(BD) theory of gravity \cite{bd} (which is one of the simplest extensions/modifications of Einstein's general
relativity) as it involves probably the simplest form of such non-linear kinetic term for the BD scalar
field. Besides, the BD scalar field (and the BD theory itself) is not of quantum origin. Rather it is
classical in nature and hence can be expected to serve as a very relevant candidate to play some role in 
the late-time evolution of the universe such as the present epoch.
Thus in this work, we would like to take the BD theory of gravity as a possible theory of
k-essence coupled to gravity and explore the role played by the BD scalar field in relation to
the unified model for dark matter and dark energy.
Therefore in our present study, the BD theory is viewed as a theory of
{\it k-essence field} (with non-canonical kinetic term) coupled {\it non-minimally} to gravity.
Namely in this context, unlike the ``scalar-tensor theory'' spirit of the original BD gravity, 
the BD scalar field is not viewed as a (scalar) part of the gravitational degrees of freedom but
instead is thought of as playing the role of a k-essence which is a matter degree of freedom.

\begin{center}
{\rm\bf II. Brans-Dicke scalar field as a unique k-essence}
\end{center}

The spirit of Brans-Dicke extension of general relativity is an attempt to properly incorporate both the
Mach's principle \cite{weinberg} and the Dirac's large number hypothesis \cite{weinberg} in which Newton's
constant is allowed to vary with space and time.
In general, the Brans-Dicke theory of gravity, in the presencs of matter with Lagrangian 
$\mathcal{L}_{M}$, is described by the action 
\begin{eqnarray}
S = \int d^4x \sqrt{g}\left[{1\over 16\pi}\left(\Phi R - \omega {{\nabla_{\alpha}\Phi
\nabla^{\alpha}\Phi }\over \Phi}\right) + \mathcal{L}_{M}\right]
\end{eqnarray}
where $\Phi $ is the BD scalar field representing the inverse of Newton's constant which is allowed to
vary with space and time and $\omega $ is the generic dimensionless parameter of the
theory. Extremizing this action then with respect to the metric $g_{\mu \nu}$ and the
BD scalar field $\Phi $ yields the classical field equations given respectively by 
\begin{eqnarray}
G_{\mu \nu} &=& R_{\mu \nu} - {1\over 2}g_{\mu \nu}R = 8\pi T^{BD}_{\mu \nu}
+ {8\pi \over \Phi}T^{M}_{\mu \nu}, \\
{\rm where} \nonumber \\
T^{BD}_{\mu \nu} &=& {1\over 8\pi}\left[{\omega \over \Phi^2}(\nabla_{\mu}\Phi \nabla_{\nu}\Phi
- {1\over 2}g_{\mu \nu}\nabla_{\alpha}\Phi \nabla^{\alpha}\Phi) + {1\over \Phi}(\nabla_{\mu}
\nabla_{\nu}\Phi - g_{\mu \nu}\nabla_{\alpha}\nabla^{\alpha}\Phi)\right], \nonumber \\
T^{M}_{\mu \nu} &=& P_{M}g_{\mu\nu} + (\rho_{M}+P_{M})U_{\mu}U_{\nu} \nonumber 
\end{eqnarray}
and
\begin{eqnarray}
\nabla_{\alpha}\nabla^{\alpha}\Phi = {8\pi \over (2\omega + 3)}T^{M\lambda}_{\lambda} 
= {8\pi \over (2\omega + 3)}(3P_{M}-\rho_{M}).
\end{eqnarray}
Note that $T^{M}_{\mu\nu}=(2/\sqrt{g})\delta (\sqrt{g}\mathcal{L}_{M})/\delta g^{\mu\nu}$, but
$T^{BD}_{\mu\nu}$ cannot be defined in a similar manner. This is due to the presence of the 
non-minimal coupling term $\sim \sqrt{g}\Phi R$ in the action which obscures the separation of
the scalar ($\Phi $) Lagrangian from the tensor ($g_{\mu\nu}$) Lagrangian. Indeed, the expression for
$T^{BD}_{\mu\nu}$ and the coefficient factor $(8\pi /\Phi )$ in front of $T^{M}_{\mu\nu}$ originate
from this non-minimal coupling term $\sim \sqrt{g}\Phi R$ upon extremizing the total action $S$ in
eq.(1) with respect to $g_{\mu\nu}$. 
Next, $P_{M}$ and $\rho_{M}$ are the pressure and the energy density of matter consisting of those of 
radiation (denoted by $rad $) and dust (denoted by $m$), i.e., 
$P_{M}=P_{rad}+P_{m}=P_{rad}$ and $\rho_{M}=\rho_{rad}+\rho_{m}$ respectively. In this perfect
fluid approximation for matter, $U^{\alpha}=dX^{\alpha}/d\tau $ (where $\tau $ denotes the proper
time) is defined to be the 4-velocity of a perfect fluid element normalized such that
$U^{\alpha}U_{\alpha}=-1$. Note that the Einstein's general relativity is
the $\omega \rightarrow \infty$ limit of this BD theory. Note also that in the action and hence in the
classical field equations, there are no {\it direct} interactions between the BD scalar field
$\Phi$ and the ordinary matter. Indeed this is the 
essential feature of the BD scalar field $\Phi$ that distinguishes it from ``dilaton'' fields
in other scalar-tensor theories such as Kaluza-Klein theories or low-energy effective string
theories where the dilaton-matter couplings generically occur as a result of dimensional reduction.
(Here we would like to stress that we shall work in the context of original BD theory format not
some conformal transformation of it. That is, we shall work in the Jordan frame, not in the Einstein frame.)
As a matter of fact, it is the original spirit \cite{bd} of BD theory of gravity in which the BD scalar
field $\Phi$ is prescribed to remain strictly massless by forbidding its direct interaction
with matter fields.  
In what follows, however, we shall first maintain the BD scalar field alone and drop 
all the others (radiation + matter) in order to uncover the role of the BD scalar field particularly in
connection with the dark matter - dark energy mixture at the present stage of the universe evolution.
In this context, the theory can now be viewed as that of a exotic
scalar matter field coupled to gravity but not a {\it pure} BD gravity theory in which both the scalar
(the BD scalar field $\Phi $) and the tensor (the metric $g_{\mu\nu}$) fields belong to the gravitational
degrees of freedom. Then one consequence of this is the fact that the statement such as ``$\omega \to \infty$
amounts to the Einstein gravity limit'' loses meaning and the BD parameter $\omega $ remains a (yet)
completely undetermined parameter of the theory. Later, however, we shall see that this BD parameter 
$\omega $ will be determined once we impose the energy-momentum conservation $\nabla_{\nu}T^{\mu\nu}=0$ 
which is nothing but the Bianchi identity, i.e., a consistency condition. \\
We now work in the spatially-flat ($k=0$) Friedman-Robertson-Walker (FRW) metric by assuming the
homogeneity and isotropy of the spacetime, i.e.,
\begin{eqnarray}
ds^2 &=& -dt^2 + a^2(t)[\frac{dr^2}{1-kr^2} + r^2(d\theta^2 + \sin^2 \theta d\phi^2)], \nonumber \\
\Phi &=& \Phi(t)
\end{eqnarray}
where $k=0$ for the spatially-flat case and $a(t)$ denotes the scale factor. Then in terms of this
FRW metric, the Friedmann equation representing the Einstein equation and the Euler-Lagrange's equation
of motion for the BD scalar field read, respectively
\begin{eqnarray}
\left(\frac{\dot{a}}{a}\right)^2 &=& \frac{\omega}{6}\left(\frac{\dot{\Phi}}{\Phi}\right)^2
- \left(\frac{\dot{a}}{a}\right)\left(\frac{\dot{\Phi}}{\Phi}\right), \\
\ddot{\Phi} &+& 3\left(\frac{\dot{a}}{a}\right)\dot{\Phi} = 0.
\end{eqnarray}
Thus in order to solve these coupled non-linear field equations, we start by assuming that
the solution ansatz is given by
\begin{eqnarray}
\Phi(t) = \frac{1}{G_{0}}a^{n}(t), ~~~a(t) = a(0)\left[1 + \frac{\beta}{\sqrt{G_{0}}}t\right]^{\alpha}
\end{eqnarray}
where $G_{0}$ denotes the value of Newton's constant at $t=0$ from which the {\it effective} Newton's 
constant, i.e., the inverse of the BD scalar field deviates and $\beta$ is some dimensionless parameter.
Note that (neglecting all matter contents) $G_{0}$ is the only fundamental scale in this theory.
First, substituting the first of this solution ansatz into the Friedmann equation in (5) yields the 
algebraic equation for the power index $n$ as $\omega n^2 - 6n - 6 =0$ which in turn gives
$n = (3 \pm \sqrt{6\omega + 9})/\omega $. Next, substituting the same ansatz this time into the BD scalar
field equation in (6) leads to a non-linear equation for only the scale factor
\begin{eqnarray}
\left(\frac{\ddot{a}}{a}\right) + (n+2)\left(\frac{\dot{a}}{a}\right)^2 = 0. \nonumber
\end{eqnarray}
Finally, substituting the second of the solution ansatz into this equation allows us to determine the
power $\alpha $ as $\alpha = 1/(n+3)$. Therefore, a solution (which indeed turns out to be a general 
solution) to the coupled Friedmann-BD scalar field equation reads
\begin{eqnarray}
a(t) &=& a(0)\left[1 + \frac{\beta}{\sqrt{G_{0}}}t\right]^{\frac{1}{n+3}}, \\
\Phi(t) &=& \frac{1}{G_{0}}a^{n}(t) = \frac{1}{G_{0}}a^{n}(0)
\left[1 + \frac{\beta}{\sqrt{G_{0}}}t\right]^{\frac{n}{n+3}} \nonumber
\end{eqnarray}
with $n = (3 \pm \sqrt{6\omega + 9})/\omega $ as obtained above. Note, here, that if we demand the
expanding universe condition, the value of $\omega$ and hence that of $n$ should be such that $(n+3)>0$. 
As we shall see later on, this condition will indeed be met in the final form of the solution.   \\
Note that in addition, the field equations in eqs.(5) and (6) should be supplemented by the 
energy-momentum conservation, $\nabla_{\nu}T^{\mu\nu}_{BD} = 0$. 
Generally speaking, if we ignore the inhomogeneities arising from the (linear) field perturbations,
the BD scalar field can be treated as a perfect fluid with stress tensor
$\tilde{T}^{BD}_{\mu\nu} = T^{BD}_{\mu\nu}/G_{0} = P_{BD}g_{\mu\nu} + (\rho_{BD}+P_{BD})U_{\mu}U_{\nu}$.
Then, $\nabla_{\nu}T^{\mu\nu}_{BD} = 0$ is given, in terms of the FRW-metric, by
\begin{eqnarray}
\dot{\rho}_{BD} = -3\left(\frac{\dot{a}}{a}\right)(\rho_{BD}+P_{BD}).
\end{eqnarray}
Note also that this energy-momentum conservation is indeed a consistency condition originating from
the (geometric) Bianchi identity $\nabla_{\nu}G^{\mu\nu}=0$ with $G_{\mu\nu}$ being the Einstein 
tensor given above in eq.(2). 
Therefore, we now need to check if the general solution in eq.(8) satisfies the energy conservation
equation given in eq.(9), which is indeed a consistency equation. As we shall see in a moment, this
energy conservation equation allows us to determine the value of the BD $\omega $-parameter to a
particular one. Thus to this end, we begin with the evaluation of the energy density and the pressure
of the BD scalar field. Thus from  
$\tilde{T}^{BD ~\mu}_{\nu} = P_{BD}\delta^{\mu}_{\nu} + (\rho_{BD}+P_{BD})U^{\mu}U_{\nu}$,
we can read off the energy density and the pressure as
\begin{eqnarray}
\rho_{BD} = - \tilde{T}^{t}_{t} = \tilde{T}_{tt},  
~~~P_{BD} = \tilde{T}^{r}_{r} = \tilde{T}^{\theta}_{\theta} = \tilde{T}^{\phi}_{\phi}.
\end{eqnarray}
Note also that the pressure can be calculated using an alternative expression,
$P_{BD} = (\tilde{T}^{\lambda}_{\lambda} + \tilde{T}_{tt})/3$. Thus using eq.(10) and the expression for the energy-momentum 
tensor for the BD scalar field given in eq.(2), we get
\begin{eqnarray}
\rho_{BD} &=& \frac{1}{16\pi G_{0}}\left[\omega \left(\frac{\dot{\Phi}}{\Phi}\right)^2 - 6 \left(\frac{\dot{a}}{a}\right)
\left(\frac{\dot{\Phi}}{\Phi}\right)\right], \\
P_{BD} &=& \frac{1}{16\pi G_{0}}\left[\omega \left(\frac{\dot{\Phi}}{\Phi}\right)^2 + 4 \left(\frac{\dot{a}}{a}\right)
\left(\frac{\dot{\Phi}}{\Phi}\right) + 2\left(\frac{\ddot{\Phi}}{\Phi}\right)\right]. \nonumber
\end{eqnarray}
Next, from the expressions for exact solutions eq.(8), it follows that
\begin{eqnarray}
\left(\frac{\dot{\Phi}}{\Phi}\right) &=& n\left(\frac{\dot{a}}{a}\right) =
\left(\frac{n}{n+3}\right)\tilde{\beta}\left[\frac{a(t)}{a(0)}\right]^{-(n+3)}, \\
\left(\frac{\ddot{\Phi}}{\Phi}\right) &=&  n\left[\left(\frac{\ddot{a}}{a}\right) + (n-1)
\left(\frac{\dot{a}}{a}\right)^2\right] = -\frac{3n}{(n+3)^2}
\tilde{\beta}^2\left[\frac{a(t)}{a(0)}\right]^{-2(n+3)} \nonumber
\end{eqnarray}
where $\tilde{\beta} = \beta /\sqrt{G_{0}}$.
Therefore from eqs.(9), (11) and (12), the energy-momentum conservation equation becomes
\begin{eqnarray}
\frac{1}{8\pi G_{0}}\left(\frac{\omega n-6}{n}\right)\left(\frac{\dot{\Phi}}{\Phi}\right)
\left[\left(\frac{\ddot{\Phi}}{\Phi}\right) - \left(\frac{\dot{\Phi}}{\Phi}\right)^2\right]
= - \frac{1}{8\pi G_{0}}\left(\frac{3}{n}\right)\left(\frac{\dot{\Phi}}{\Phi}\right)
\left[\left(\frac{\ddot{\Phi}}{\Phi}\right) + \left(\frac{\omega n-1}{n}\right)
\left(\frac{\dot{\Phi}}{\Phi}\right)^2\right] \nonumber
\end{eqnarray}
which, in turn, gives
\begin{eqnarray}
\omega = -\frac{3}{2}, ~~~n = - 2
\end{eqnarray}
and hence finally the solution reads
\begin{eqnarray}
a(t) = a(0)\left[1 + \frac{\beta}{\sqrt{G_{0}}}t\right], 
~~~\Phi(t) = \frac{1}{G_{0}}a^{-2}(0)
\left[1 + \frac{\beta}{\sqrt{G_{0}}}t\right]^{-2}. 
\end{eqnarray}
Note also that this set of values in eq.(13) is consistent with 
$n = (3 \pm \sqrt{6\omega + 9})/\omega $ that has been determined from the field equations. 
As mentioned earlier, the field equations leave the BD 
$\omega $-parameter and the power index $n$ undetermined but it is the energy-momentum conservation (which
is a consistency condition) that finally determines their values. Having obtained the explicit expressions
for energy density and pressure of the BD scalar field (being treated as a perfect fluid), we now
attempt to determine the equation of state and the speed of sound. Firstly, using eqs.(11), (12) and (13),
we have
\begin{eqnarray}
\rho_{BD} = \frac{1}{16\pi G_{0}}\left[6\tilde{\beta}^2 \left(\frac{a(t)}{a(0)}\right)^{-2}\right],
~~~P_{BD} = - \frac{1}{16\pi G_{0}}\left[2\tilde{\beta}^2 \left(\frac{a(t)}{a(0)}\right)^{-2}\right]
\end{eqnarray}
which yields the equation of state to be 
$w_{BD} = P_{BD}/\rho_{BD} = -2\tilde{\beta}^2/6\tilde{\beta}^2 = -1/3$ 
(and hence, for later use, note that $(\rho_{BD} + 3P_{BD}) = 0$), namely,
\begin{eqnarray}
P_{BD} = -\frac{1}{3}\rho_{BD}
\end{eqnarray}
meaning that the BD scalar field turns out to be a {\it barotropic} (i.e., constant $w$) perfect fluid.
Next, the computation of the sound speed needs more careful treatment. Namely, using again eqs.(11) and (12),
it follows that
\begin{eqnarray}
\dot{\rho}_{BD} &=& \frac{1}{8\pi G_{0}}\left(\frac{3}{2}\right)\left(\frac{\dot{\Phi}}{\Phi}\right)
\left[\left(\frac{\ddot{\Phi}}{\Phi}\right) - \left(\frac{\dot{\Phi}}{\Phi}\right)^2\right] =
-\frac{1}{8\pi G_{0}}\left[6\tilde{\beta}^3\left(\frac{a(t)}{a(0)}\right)^{-3}\right], \\
\dot{P}_{BD} &=& \frac{1}{8\pi G_{0}}\left(\frac{1}{2}\right)\left[7\left(\frac{\dot{\Phi}}{\Phi}\right)^3 -
9\left(\frac{\dot{\Phi}\ddot{\Phi}}{\Phi^2}\right) + 2\left(\frac{\dddot{\Phi}}{\Phi}\right)\right] =
\frac{1}{8\pi G_{0}}\left[2\tilde{\beta}^3\left(\frac{a(t)}{a(0)}\right)^{-3}\right]. \nonumber
\end{eqnarray}
Therefore, the speed of sound in this BD scalar field fluid turns out to be
\begin{eqnarray}
c^2_{s} = \frac{dP_{BD}}{d\rho_{BD}} = \frac{dP_{BD}/dt}{d\rho_{BD}/dt} = \frac{2\tilde{\beta}^3}
{-6\tilde{\beta}^3} = -\frac{1}{3}
\end{eqnarray}
where we followed the spirit of definition for the effective sound speed of perturbations 
suggested by Ratra \cite{ratra}. Namely, it is {\it negative}. 
Indeed, the negative definite sound speed squared generically signals the instabilities. 
Types of ordinary matter such as radiation or dust exhibit non-negative pressure and hence non-negative
sound speed squared. But for rather an exotic type of matter like the k-essence, it is not so surprising to
have negative sound speed squared (as long as it is a barotropic fluid) as it may well result from the 
negative pressure. Indeed, $c^2_{s}$ in our BD scalar field fluid is essentially the same as the equation 
of state $w_{BD}$ and hence is negative. 
We shall now be more specific in relating the negative definite sound speed squared to the instabilities
of perturbations. Indeed, this issue of instability plays a key role in the so-called XCDM model for
dark energy \cite{peebles2} as well. And to this end, we shall follow the general formulation given by 
Peebles and Ratra \cite{peebles2} and apply it to the present model of BD scalar field fluid.
In the context of {\it linear} perturbation theory in which the density perturbation is described by
$\rho_{BD}(t, \vec{x}) = <\rho_{BD}(t)> + \delta \rho_{BD}(t, \vec{x})$ 
(where $<...>$ denotes the background (mean) value), the equation of energy-momentum conservation, 
$\nabla_{\nu}T^{\mu\nu}_{BD} = 0$ yields
$\delta \ddot{\rho}_{BD} = c^2_{s}\nabla^2 \delta \rho_{BD}$ where $c^2_{s} = dP_{BD}/d\rho_{BD}$ as usual. \\
(1) If $c^2_{s}>0$, this perturbation equation becomes a wave equation whose solution would be given, say,
by $\delta \rho_{BD} = \delta \rho_{0}e^{-i\omega t + i\vec{k}\cdot \vec{x}}$ and the usual positive
definite sound speed squared (or the real value of sound speed) indicates the propagating mode of 
perturbation. \\
(2) But if $c^2_{s}<0$, the solution to this perturbation equation would be given by
$\delta \rho_{BD} = \delta \rho_{0}e^{\pm |\omega| t + i\vec{k}\cdot \vec{x}}$ and the negative 
definite sound speed squared (or the imaginary value of sound speed) indiates the (exponentially) growing 
or decaying mode signaling the instability of perturbation. Besides, since 
$c^2_{s} = dP_{BD}/d\rho_{BD} = - |\omega|^2/|\vec{k}|^2 < 0$, this ``imaginary'' sound speed indicates
that the increase (decrease) in density leads to lowering (growth) of pressure supporting the emergence
of instability. \\
Some ways to avoid the negative definite sound speed squared and hence the unstable inhomogeneities
when encountered have been discussed in the literature \cite{sound}. 
Among them, one very convincing argument is due to Steinhardt \cite{steinhardt}.
It suggests that although the imaginary sound speed might mean accelerated collapse of inhomogeneities,
such instability could be avoided at least at the subhorizon scale by taking into account the dependence
of the sound speed on the wavelength characterizing the instabilities.  \\
Thus far, we have determined the behaviors 
\begin{eqnarray}
a(t) \sim t, ~~~\rho_{BD} \sim \frac{1}{a^{2}(t)}, ~~~P_{BD} = -\frac{1}{3}\rho_{BD}.
\end{eqnarray}
Based on this observation, we now discuss the role played by the existence of this scalar field dominated
era in relation to the so-called ``coincidence problem'' asking why the transition from the matter-dominated
era to the (observed) cosmic acceleration phase occurs only at the present stage.
First, regardless of the type of matter (i.e., whether
it is the ordinary ones like radiation and dust or the k-essence such as the BD scalar field), one can easily
realize that the strong energy condition can be represented by
\begin{eqnarray}
R_{\mu\nu}\xi^{\mu}\xi^{\nu} &=& 8\pi G_{0}[T_{\mu\nu} - \frac{1}{2}g_{\mu\nu}T^{\lambda}_{\lambda}]
\xi^{\mu}\xi^{\nu} \nonumber \\
&=& 4\pi G_{0}[3P + \rho ] \geq 0 
\end{eqnarray}
where $\xi^{\mu}$ denotes the future-directed timelike unit normal vector and we used
$\rho = T_{\mu\nu}\xi^{\mu}\xi^{\nu}$ and $T^{\lambda}_{\lambda} = (3P - \rho )$.
Next, in the standard cosmology in which ordinary type of matter (radiation or dust) is coupled to
Einstein gravity, the time-time component of the Einstein field equation gives
\begin{eqnarray}
\left(\frac{\ddot{a}}{a}\right) = - \frac{4\pi G_{0}}{3}(3P + \rho )
\end{eqnarray}
whereas in the present case of k-essence model in which the BD scalar field plays the role of the
k-essence, the time-time component of the metric field equation reads
\begin{eqnarray}
\left(\frac{\ddot{a}}{a}\right) = - \frac{\omega }{3}\left(\frac{\dot{\Phi}}{\Phi}\right)^2 -
\frac{1}{3}\left(\frac{\ddot{\Phi}}{\Phi}\right).
\end{eqnarray}
Now from eq.(11) above, we have
\begin{eqnarray}
(3P_{BD} + \rho_{BD}) = \frac{1}{16\pi G_{0}}\left[4\omega \left(\frac{\dot{\Phi}}{\Phi}\right)^2 +
6\left(\frac{\dot{a}}{a}\right)\left(\frac{\dot{\Phi}}{\Phi}\right) + 
6\left(\frac{\ddot{\Phi}}{\Phi}\right)\right].
\end{eqnarray}
Then using eq.(12) and by substituting the set of values $\omega = -3/2$, $n= - 2$ determined above,
the eqs.(22) and (23) yield
\begin{eqnarray}
\left(\frac{\ddot{a}}{a}\right) = - \frac{8\pi G_{0}}{9}(3P_{BD} + \rho_{BD} ).
\end{eqnarray}
Thus to summarize, for both the ordinary matter case and the k-essence case, the strong energy condition is always
represented by the factor $(3P + \rho )$ and it is its sign which always determines whether the universe 
decelerates or accelerates. With this preparation, we now recall the physical quantities characterizing
each era in the universe evolution which are summarized in TABLE I. 
\begin{table}
\centering
\begin{tabular}{|c|c|c|} \hline
$\rho_{rad}\sim \frac{1}{a^{4}(t)}$, $P_{rad}=\frac{1}{3}\rho_{rad}$ & 
$a(t) \sim t^{1/2}$, $(3P_{rad}+\rho_{rad})>0$ & deceleration  \\ \hline
$\rho_{m}\sim \frac{1}{a^{3}(t)}$, $P_{m}=0$ & 
$a(t) \sim t^{2/3}$, $(3P_{m}+\rho_{m})>0$ & deceleration  \\ \hline
$\rho_{BD}\sim \frac{1}{a^{2}(t)}$, $P_{BD}=-\frac{1}{3}\rho_{BD}$ & 
$a(t) \sim t $, $(3P_{BD}+\rho_{BD})=0$ & zero acceleration  \\ \hline
$\rho_{acc}=const.$, $P_{acc}= w\rho_{acc}$  $(-1\leq w< -1/3)$ & 
$a(t) \sim t^{m}$ $(m>1)$, $(3P_{acc}+\rho_{acc})<0$ & acceleration  \\ \hline
\end{tabular}
\caption{Summary of cosmic evolution in the presence of the scalar field-dominated era.} \label{t1}
\end{table}
The four rows represent radiation-dominated, matter(dust)-dominated, (BD) scalar field-dominated and 
the accelerating eras, respectively. Of course, the {\it total} energy density $\rho$ and the pressure
$P$ consist of all the contributions coming from each component, radiation, dust, BD scalar field and
some unknown entity (denoted by the subscript ``$acc$'') leading to the acceleration, namely
\begin{eqnarray}
\rho &=& \rho_{rad} + \rho_{m} + \rho_{BD} + \rho_{acc}, \\
P &=& P_{rad} + P_{m} + P_{BD} + P_{acc} \nonumber
\end{eqnarray}
with $P_{m}=0$ as it is approximated as a dust. 
The presence of the last (i.e., the most recent) era, namely the acceleration phase has been
introduced based on the cosmological observation of the present large scale structure such as the 
anisotropy in CMBR and the luminosity of type Ia supernovae at high redshift which all suggest that
the universe is currently undergoing cosmic acceleration and is dominated by dark energy component
with negative pressure \cite{science}. And we placed the (BD) scalar field-dominated era in between
the matter-dominated and the accelerating eras based on the way $\rho_{BD}$ scales with
the scale factor $a(t)$, namely $\rho_{BD} \sim 1/a^2(t)$. Thus from the scaling behavior of the energy
density of each epoch, it is straightforward to see that as the universe expands (i.e., as $a(t)$
grows), $\rho_{rad}$ rises and falls first, $\rho_{m}$ next, then does $\rho_{BD}$ and lastly
$\rho_{acc}$ begins to dominate. Of course, the behavior of other physical 
quantities during the (BD) scalar field-dominated epoch also strongly suggests that it should be inserted 
between the matter-dominated and the accelerating eras. Thus with this new era being inserted, it is 
interesting to realize that now we are witnessing a grand picture of universe evolution in which,
as the universe expands, or as time goes on, the energy density $\rho $ dilutes more and more slowly, 
the pressure $P$ keeps decreasing from positive value eventually toward negative one, the scale factor
grows more and more rapidly and lastly, the strong energy condition (i.e., the sign of 
$(3P+\rho )$) moves from ``yes'' towards ``no'' leading to the transit of cosmic evolution from
{\it deceleration} towards {\it acceleration} past zero acceleration during the newly-inserted
(BD) scalar field-dominated era.  Among others, therefore, it appears that the newly found presence of
the (BD) scalar field-dominated epoch provides a picture of smooth transition from the decelerating 
matter-dominated era to the epoch of current acceleration as the acceleration is {\it zero} during this
era. Moreover, the BD scalar field itself is known to possess a generic {\it dark} nature as it is not
allowed to have direct interactions (couplings) with ordinary matter (radiation + dust) from the outset,
namely at the Largangian level since otherwise it would violate the cherished equivalence principle
as has been originally pointed out by Brans and Dicke themselves. \\ 
Indeed, one of the features of the present
study (in which the BD scalar field is taken to play the role of a k-essence field) that distinguishes
it from other k-essence models can be summarized as follows. 
It provides no {\it direct} mechanism for the arrival (or emergence) of the present cosmic acceleration 
phase. Instead, it presents a good reason or evidence explaining why the matter-dominated era is to be 
followed by an accelerating phase as the acceleration is ``zero'' during the BD scalar field-dominated
era in between the two. 
Namely, it provides a crossing bridge between the matter-dominated and the
accelerating eras that has been missing from the scene thus far. From the theoretical perspective,
this is certainly a physically natural and meaningful prediction of our model for dark energy. 
On the observational side, however, there has been no evidence for such zero acceleration epoch to date.

\begin{center}
{\rm\bf III. Effects of BD scalar field (a k-essence) on the late-time universe evolution}
\end{center}

Thus far we have ignored the possible contributions from other types of matter and concentrated on the
role played by the BD scalar field, i.e., the k-essence in order particularly to explore the (BD) scalar
field-dominated era when the contribution from the k-essence overwhelms all others. And as a result, we
found that such (BD) scalar field-dominated era is a yet-unknown {\it zero acceleration} epoch that
should be inserted in between the decelerating matter-dominated and the accelerating eras acting as a
``crossing bridge'' between the two. Upon realizing this, then, our next mission should be a closer 
study of the effects of the k-essence (i.e., the BD scalar
field) on the evolutionary behavior of the matter-dominated and accelerating eras. Thus we now should
include all the three components, the dust-like matter, BD scalar field and the entity that drives the
current cosmic acceleration. For the sake of definiteness of our demonstration, this entity responsible
for the cosmic acceleration shall be taken as the cosmological constant $\Lambda $. Therefore,
this study of the effects of BD scalar field on the nature of matter-dominated and accelerating eras
can be thought of as our attempt to build a ``unified model'' for dark matter-dark energy that is
currently under intensive exploration in the theoretical cosmology. \\
We now reconsider the action for the Brans-Dicke theory of gravity, this time in the presence of the
cosmological constant $\Lambda $ with mass dimension 4 as well as the dust-like matter (denoted by $m$), 
\begin{eqnarray}
S = \int d^4x \sqrt{g}\left[{1\over 16\pi}\left(\Phi R - \omega {{\nabla_{\alpha}\Phi
\nabla^{\alpha}\Phi }\over \Phi}\right) - \Lambda + \mathcal{L}_{m}\right].
\end{eqnarray}
Again extremizing this action with respect to the metric $g_{\mu \nu}$ and the
BD scalar field $\Phi $ yields the classical field equations given respectively by 
\begin{eqnarray}
G_{\mu \nu} &=& R_{\mu \nu} - {1\over 2}g_{\mu \nu}R + {8\pi \over \Phi}\Lambda g_{\mu \nu} = 
8\pi T^{BD}_{\mu \nu} + {8\pi \over \Phi}T^{m}_{\mu \nu}, \\
{\rm where} \nonumber \\
\tilde{T}^{BD}_{\mu \nu} &=& T^{BD}_{\mu\nu}/G_{0} = P_{BD}g_{\mu\nu} + (\rho_{BD}+P_{BD})U_{\mu}U_{\nu}, 
\nonumber \\
T^{m}_{\mu \nu} &=& P_{m}g_{\mu\nu} + (\rho_{m}+P_{m})U_{\mu}U_{\nu} \nonumber 
\end{eqnarray}
and
\begin{eqnarray}
\nabla_{\alpha}\nabla^{\alpha}\Phi = {8\pi \over (2\omega + 3)}\left[T^{m\lambda}_{\lambda} - 4\Lambda \right] 
= {8\pi \over (2\omega + 3)}\left[(3P_{m}-\rho_{m}) - 4\Lambda \right]
\end{eqnarray}
with $P_{m} = 0$ in the dust approximation of matter. As usual we work in the spatially-flat FRW metric
given in eq.(4) in terms of which the Friedmann equation representing the Einstein equation and 
the Euler-Lagrange's equation of motion for the BD scalar field are given, respectively
\begin{eqnarray}
\left(\frac{\dot{a}}{a}\right)^2 &=& \frac{8\pi }{3\Phi }(\rho_{m}+\Lambda ) 
+ \frac{\omega}{6}\left(\frac{\dot{\Phi}}{\Phi}\right)^2
- \left(\frac{\dot{a}}{a}\right)\left(\frac{\dot{\Phi}}{\Phi}\right), \\
\ddot{\Phi} &+& 3\left(\frac{\dot{a}}{a}\right)\dot{\Phi} = \frac{8\pi }{(2\omega +3)}
\left[(\rho_{m}-3P_{m}) + 4\Lambda \right].
\end{eqnarray}
Once again, we stress that these field equations have to be supplemented by the consistency conditions.
Namely, in addition to these classical field equations for the metric $g_{\mu\nu}$ and the BD scalar
field $\Phi $, there is one more set of equations which are consistency conditions as they result from
the {\it geometric} Bianchi identity $\nabla_{\nu}G^{\mu\nu} = 0$. Thus from eq.(27) we have
\begin{eqnarray}
0 = \nabla_{\nu}(R^{\mu \nu} - {1\over 2}g^{\mu \nu}R) = \nabla_{\nu}
\left[-{8\pi \over \Phi}\Lambda g^{\mu \nu} + {8\pi \over \Phi}T^{\mu \nu}_{m} + 8\pi T^{\mu \nu}_{BD}\right].
\end{eqnarray} 
Note first that, according to the original spirit of Brans and Dicke \cite{bd}, in order not to interfere
with the successes of the equivalence principle, the BD scalar field $\Phi $ is assumed not to enter
into the equations of motion of ordinary (dust) matter so that $T^{\mu\nu}_{m}$ obeys
the usual conservation law $\nabla_{\nu}T^{\mu\nu}_{m}=0$ which, in terms of the FRW metric, takes the
familiar form
\begin{eqnarray}
\dot{\rho}_{m} = -3\left(\frac{\dot{a}}{a}\right)(\rho_{m}+P_{m})
\end{eqnarray}
where again $P_{m}=0$ in dust approximation. Therefore, we are left with
\begin{eqnarray}
0 = 8\pi \Lambda \frac{(\partial_{\nu}\Phi)}{\Phi^2}g^{\mu\nu} - 8\pi \frac{(\partial_{\nu}\Phi)}{\Phi^2}T^{\mu\nu}_{m}
+ 8\pi G_{0}\nabla_{\nu}[P_{BD}g^{\mu\nu} + (\rho_{BD}+P_{BD})U^{\mu}U^{\nu}]
\end{eqnarray}
which, again in terms of the FRW metric, becomes
\begin{eqnarray}
\dot{\rho}_{BD} + 3\left(\frac{\dot{a}}{a}\right)(\rho_{BD}+P_{BD}) = \frac{1}{G_{0}}
(\rho_{m}+\Lambda )\left(\frac{\dot{\Phi}}{\Phi^2}\right).
\end{eqnarray}
Thus in order to determine the natute of late-time universe evolution, one in principle has to attempt to 
solve the coupled, non-linear field equations given in eqs.(29) and (30) subject to the consistency 
conditions in eqs.(32) and (34). However, one cannot solve them with both the 
dust-like matter and the cosmological constant present as they carry distinct equations of state.
In practice, therefore, we consider the two stages ; first, the matter-to-scalar field-dominated 
era (MATTER-TO-SCALAR) transition period and second, the scalar field-dominated-to-accelerating 
phase (SCALAR-TO-ACCELERATION) transition period.  
\\
{\bf 1. MATTER-TO-SCALAR transition period}
\\
At this stage, the energy density of universe is assumed to be dominated by those of dust-like matter,
$\rho_{m}$ and BD scalar field, $\rho_{BD}$. First of all, in order to solve the coupled Friedmann 
equation and the BD scalar field equation in eqs.(29) and (30) (with $\Lambda $ absent for the case at hand), 
the dependence of the matter energy density $\rho_{m}$ on the scale factor has to be given. As usual, this
can be determined using the equation of state for the matter-dominated era, $P_{m}=0$ and the
energy-momentum conservation equation (32) which is one of the consistency equations. The result is the
familiar dependence, $\rho_{m}\sim 1/a^3(t)$, or more precisely,
\begin{eqnarray}
\rho_{m} = \frac{1}{G^2_{0}}\left[\frac{a(t)}{a(0)}\right]^{-3}.
\end{eqnarray}
Then the coupled field equations for this transition period becomes
\begin{eqnarray}
\left(\frac{\dot{a}}{a}\right)^2 &=& \frac{8\pi }{3G^2_{0} }\frac{\kappa}{\Phi a^3} 
+ \frac{\omega}{6}\left(\frac{\dot{\Phi}}{\Phi}\right)^2
- \left(\frac{\dot{a}}{a}\right)\left(\frac{\dot{\Phi}}{\Phi}\right), \\
\ddot{\Phi} &+& 3\left(\frac{\dot{a}}{a}\right)\dot{\Phi} = \frac{8\pi \kappa}{(2\omega +3)G^2_{0}}
\frac{1}{a^3}
\end{eqnarray}
where $\kappa \equiv a^3(0)$
and of course these field equations have to be supplemented by the remaining consistency equation
\begin{eqnarray}
\dot{\rho}_{BD} + 3\left(\frac{\dot{a}}{a}\right)(\rho_{BD}+P_{BD}) = \frac{\kappa}{G^3_{0}}
\frac{1}{a^3}\left(\frac{\dot{\Phi}}{\Phi^2}\right).
\end{eqnarray}
We now start with the solution ansatz
\begin{eqnarray}
\Phi(t) = \frac{1}{G_{0}}\left[1 + \chi t\right]^{\gamma}, 
~~~a(t) = a(0)\left[1 + \chi t\right]^{\alpha}.
\end{eqnarray}
Substituting these solution ansatz into the coupled field equations (36) and (37) above yields,
\begin{eqnarray}
\alpha = \frac{2(\omega +1)}{(3\omega +4)}, ~~~\gamma = \frac{2}{(3\omega +4)},
~~~\chi^2 = \frac{4\pi}{G_{0}}\frac{(3\omega +4)}{(2\omega +3)}.
\end{eqnarray}
Thus basically the solutions behave as
\begin{eqnarray}
\Phi(t) \sim t^{2/(3\omega +4)}, ~~~a(t) \sim t^{2(\omega +1)/(3\omega +4)}
\end{eqnarray}
and they have actually been known \cite{weinberg} for some time in the Brans-Dicke cosmology.
Note that in the standard Einstein gravity limit $\omega \to \infty$ in which the dynamics of the
BD scalar field is washed out, i.e., $\Phi \to 1/G_{0}$, one recovers the standard behavior
for the matter-dominated era, $a(t) \sim t^{2/3}$. 
With these exact solutions at hand, next we consider the behavior of the BD scalar field
(being treated as a perfect fluid) in this transition period. Recall, first that
its energy density and pressure are given by eq.(11) earlier, which have been obtained using eqs.(2)
and (10). Thus by plugging the exact solutions for the case at hand, eqs.(39) and (40) into eq.(11),
we end up with
\begin{eqnarray}
\rho_{BD} &=& -\frac{1}{G^{2}_{0}}\left[\frac{(5\omega +6)}{(2\omega +3)(3\omega +4)} 
\left(\frac{a(t)}{a(0)}\right)^{-2/\alpha}\right],  \\
P_{BD} &=& \frac{1}{G^{2}_{0}}\left[\frac{2(\omega +1)}{(2\omega +3)(3\omega +4)} 
\left(\frac{a(t)}{a(0)}\right)^{-2/\alpha}\right]  \nonumber
\end{eqnarray}
where $\alpha$ is as given in eq.(40).
Here a causion needs to be exercised. Namely, these expressions for $\rho_{BD}$ and $P_{BD}$ do not
necessarily mean that $\rho_{BD}<0$ and $P_{BD}>0$. Indeed, it is the other way around if we demand
the positive-definite energy density for the BD scalar field, i.e., $\rho_{BD}>0$. Then it immediately
follows that $P_{BD}<0$. To see this is indeed the case, note that demanding $\rho_{BD}>0$ leads to
$\omega <-3/2$ or $-4/3<\omega <-6/5$ which, in turn, indicates $(2\omega +3)(3\omega +4)>0$ and 
$(\omega +1)<0$ and also $(5\omega +6)<0$. Therefore, clearly $P_{BD}<0$. Besides, this condition
$\rho_{BD} > 0$ (particularly $\omega <-3/2$) also guarantees the condition of cosmic expansion
$\alpha = 2(\omega +1)/(3\omega +4) > 0$. Next note that as a result,
the equation of state of the BD scalar field perfect fluid in this transition period is given by
\begin{eqnarray}
w_{BD} = \frac{P_{BD}}{\rho_{BD}} = -\frac{2(\omega +1)}{(5\omega +6)} < 0.
\end{eqnarray}
Namely, it is {\it negative} implying that already in this matter-to-scalar field-dominated era transition 
period, the BD scalar field behaves as a negative pressure component. In the earlier section in which
we maintained the BD scalar field alone and ignored the contributions from other types of matter, it
has been realized that the equation of state of the BD scalar field is $w_{BD} = P_{BD}/\rho_{BD}
= -1/3$ implying the emergence of zero acceleration epoch driven by the negative pressure. This
negative pressure nature of the BD scalar field as a k-essence, therefore, appears to begin already in
this transition period. Next, the total pressure in this transition period is given by 
$P_{tot} = P_{m} + P_{BD} = P_{BD} < 0$. Thus it is the BD scalar field that renders the total pressure
at this stage negative and this, in turn, implies that the BD scalar field would therefore act to
{\it speed up} the expansion rate of the matter-dominated era. This actually will turn out to be the
case since $\alpha > 2/3$ as will be shown shortly in eq.(51) later on.
Lastly, note that these solutions in eqs.(39) and (40) indeed satisfy the remaining consistency condition
in eq.(38). It can be checked in a straightforward manner using the expressions for $\rho_{BD}$ and
$P_{BD}$ given in eq.(42) and the exact solutions in eqs.(39) and (40). This consistency condition,
however, does not determine the BD $\omega$-parameter to a particular value.
\\
{\bf 2. SCALAR-TO-ACCELERATION transition period}
\\
At this stage, the universe energy density is assumed to be dominated by those of BD scalar field, 
$\rho_{BD}$ and the cosmological constant $\Lambda $.
Particularly, the study of this second stage involving essentially the two components, 
the BD scalar field $\Phi $ and the cosmological constant $\Lambda $, can be thought of as our proposal 
for a k-essence model for dark energy. Then the difference between our dark energy model and those of other
quintessence/k-essence theories lies in the fact that here in our model, we are interested in the way
the presence of the BD scalar field, i.e., a k-essence, modifies the evolutionary behavior of the
vacuum energy ($\Lambda$)-dominated epoch (which has an exponentially expanding nature in the standard general
relativity context) while there, the quintessence/k-essence fields themselves (without the $\Lambda$-term)
are expected to generate an accelerating expansion of some sort. \\ 
The coupled Friedmann equation and the BD scalar field equation for the case at hand amount to setting 
$\rho_{m} = 0$ in eqs.(29) and (30), namely
\begin{eqnarray}
\left(\frac{\dot{a}}{a}\right)^2 &=& \frac{8\pi }{3\Phi}\Lambda 
+ \frac{\omega}{6}\left(\frac{\dot{\Phi}}{\Phi}\right)^2
- \left(\frac{\dot{a}}{a}\right)\left(\frac{\dot{\Phi}}{\Phi}\right), \\
\ddot{\Phi} &+& 3\left(\frac{\dot{a}}{a}\right)\dot{\Phi} = \frac{8\pi}{(2\omega +3)}
4\Lambda.
\end{eqnarray}
Again, these field equations should be supplemented by the remaining consistency equation
\begin{eqnarray}
\dot{\rho}_{BD} + 3\left(\frac{\dot{a}}{a}\right)(\rho_{BD}+P_{BD}) = \frac{\Lambda}{G_{0}}
\left(\frac{\dot{\Phi}}{\Phi^2}\right).
\end{eqnarray}
Again, we begin with the solution ansatz
\begin{eqnarray}
\Phi(t) = \frac{1}{G_{0}}\left[1 + \chi t\right]^{\gamma}, 
~~~a(t) = a(0)\left[1 + \chi t\right]^{\alpha}.
\end{eqnarray}
Substituting these solution ansatz into the coupled field equations (44) and (45) yields in this time,
\begin{eqnarray}
\alpha &=& \frac{(2\omega +1)}{2},  ~~~\gamma = 2, \\
\chi &=& \tilde{\chi}/h, ~~~\tilde{\chi}^2 = \frac{8\pi G_{0}}{3}\Lambda,
~~~h^2 = \frac{(2\omega +3)(6\omega +5)}{12}.  \nonumber
\end{eqnarray}
Thus the solutions behave as
\begin{eqnarray}
\Phi(t) = \frac{1}{G_{0}}\left[1 + \chi t\right]^{2}, 
~~~a(t) = a(0)\left[1 + \chi t\right]^{(2\omega +1)/2}.  \nonumber
\end{eqnarray}
Having these exact solutions with us, again we consider the behavior of the BD scalar field
(i.e., a k-essence) in this transition period.  
By plugging the exact solutions for the case at hand, eqs.(47) and (48) into eq.(11),
we are left with 
\begin{eqnarray}
\rho_{BD} &=& -\Lambda \left[\frac{4(4\omega +3)}{(2\omega +3)(6\omega +5)} 
\left(\frac{a(t)}{a(0)}\right)^{-2/\alpha}\right],  \\
P_{BD} &=& \Lambda \left[\frac{8(3\omega +2)}{(2\omega +3)(6\omega +5)} 
\left(\frac{a(t)}{a(0)}\right)^{-2/\alpha}\right] \nonumber
\end{eqnarray}
with $\alpha$ being given in eq.(48) and as a result,
the equation of state of the BD scalar field perfect fluid in this transition period is given by
\begin{eqnarray}
w_{BD} = \frac{P_{BD}}{\rho_{BD}} = -\frac{2(3\omega +2)}{(4\omega +3)} < 0.
\end{eqnarray}
In this transition period, however, the cosmic acceleration is supposed to set in. Thus one should
demand $\alpha = (2\omega +1)/2 > 1$ or $\omega > 1/2$ and this, in turn, leads to $\rho_{BD}<0$ and
$P_{BD}>0$ as $\Lambda > 0$ from eq.(48). Namely at this stage, the BD scalar field turns around and
begins to behave as a {\it positive} pressure component making the total pressure 
$P_{tot} = P_{\Lambda} + P_{BD} = -\Lambda + P_{BD}$ {\it less} negative and hence {\it slowing down}
the accelerated expansion rate from the exponential law to the power law as we shall discuss in more
detail below. Note, however, that since $a(t) \sim t^{\alpha}$ and hence 
$\rho_{BD}, P_{BD} \sim a^{-2/\alpha}(t) \sim 1/t^{2}$, the {\it positive} contribution of $P_{BD}$
to $P_{tot} = -\Lambda + P_{BD}$ would eventually become negligible as time goes on and the same
argument holds for the total energy density $\rho_{tot} = \rho_{\Lambda} + \rho_{BD} = \Lambda + \rho_{BD}$
namely, the {\it negative} contribution of $\rho_{BD}$ to $\rho_{tot}$ would become more and more
negligible. This last observation, then, appears to indicate that the accelerated expansion would
eventually ``speed up''.
Lastly, note that these solutions in eqs.(47) and (48) indeed satisfy the remaining consistency condition
in eq.(46). It can be checked straightforwardly using the expressions for $\rho_{BD}$ and
$P_{BD}$ given in eq.(49) and the exact solutions in eqs.(47) and (48). Again, this consistency condition
does not determine the BD $\omega$-parameter to a particular value.  \\
These solutions in eqs.(47) and (48) have indeed been found some time ago in the so-called 
{\it extended inflation} model \cite{la} which attempts to resolve the {\it graceful exit} problem and
hence to save the original spirit of ``old inflation'' scenario. There the punchline was the realization that
by replacing the general relativity content with the Brans-Dicke gravity, the accelerated expansion rate slows
down from the exponential expansion to the power law one while the false-to-true vacuum phase transition rate
remains the same and hence leading to the successful percolation and exit to the radiation-dominated era.
Note again that in the standard Einstein gravity limit $\omega \to \infty$ in which the dynamics of the
BD scalar field is washed out, i.e., $\Phi \to 1/G_{0}$ (as $\chi \to 0$), one recovers the familiar 
exponential expansion, $a(t)=a(0)e^{\tilde{\chi}t}$ in the Einstein gravity framework.
Then the essential difference between the extended inflation then and the late-time cosmic acceleration
now but with the same solutions would be ; (1) in the context of inflation at early universe, the model
has to be constrained by such conditions as the enough expansion to resolve the difficulties with the 
hot big-bang model like the horizon, flatness and topological defects problems and the generation of 
density perturbation to some proper degree so as not to contradict with the observed anisotropy in 
CMBR \cite{wmap}.
(2) In the context of the late-time cosmic acceleration, on the other hand, the present dark energy model
based on the BD theory should be tested carefully, say, by the current WMAP data \cite{wmap} and
the luminosity of type Ia supernova at high redshift \cite{science} both of which would provide a severe
constraint on the current cosmic acceleration rate. Then eventually this would set a constraint on the
value of the BD $\omega$-parameter at this stage through the parameters $\alpha$ and $\chi$ in eq.(48).
\\
{\bf 3. Remarks}
\\
(I) We go back to the case of MATTER-TO-SCALAR transition period studied above and examine
the nature of solutions there. First, the condition of ``cosmic expansion'' amounts to demanding
$\alpha = 2(\omega +1)/(3\omega +4) >0$, which yields $\omega <-4/3$ or $\omega >-1$. Secondly, 
there we demanded $\rho_{BD}>0$ which led to $\omega <-3/2$ or $-4/3<\omega <-6/5$.
Thus these two conditions, when combined, give $\omega <-3/2$. This, in turn, allows us to conclude that
\begin{eqnarray}
\left(\alpha - \frac{2}{3}\right) = \frac{-2}{3(3\omega +4)} > 0. 
\end{eqnarray}
This inequality indicates that the expansion rate of the matter-dominated era in the presence of the
BD scalar field, which is a k-essence, (or equivalently in the context of BD gravity theory) namely,
$a(t)\sim t^{\alpha}$, turns out to be greater than that, i.e., $a(t)\sim t^{2/3}$ in the absence of 
the k-essence field (or in the context of general relativity). 
Note also that from eqs.(35), (39) and (42), 
$\rho_{m} \sim a^{-3}(t) \sim 1/t^{3\alpha} < \rho_{BD} \sim a^{-2/\alpha}(t) \sim 1/t^2$ as 
$\alpha > 2/3$. Recall from section II that $a(t) \sim t$ and thus $\rho_{BD} \sim a^{-2}(t) \sim 1/t^2$
in the BD scalar field-dominated era. This observation indicates that as the evolution proceeds,
the BD scalar field component takes over the energy density of the matter-dominated era and as a result, 
the BD scalar field-dominated era does indeed arrive as we assumed in the section II. And this supports
our suggestion for the grand picture of universe evolution summarized in TABLE I. \\
(II) Next, in the case of SCALAR-TO-ACCELERATION transition period, we learned that the accelerated 
expansion rate in the vacuum energy ($\Lambda$)-dominated era slows down to a power law by introducing
the BD scalar field, which is again a k-essence (or equivalently in the context of BD theory). \\
Now these two observations (I) and (II) appear to imply that the effects of the presence of the BD scalar
field (playing the role of an unique k-essence) on the late-time evolution of the universe can be
summarized as follows ; the presence of the k-essence speeds up the expansion rate of the matter-dominated era
(and hence {\it reduces} the deceleration) while slows down the expansion rate of the late-time cosmic
acceleration phase (and hence again {\it reduces} the acceleration) ! 
This effect coming from the presence of the BD scalar field has been summarized in TABLE II. 
\begin{table}
\centering
\begin{tabular}{|c|c|c|} \hline
       & {\rm General Relativity (without BD scalar)} & {\rm Brans-Dicke Theory (with BD scalar)} \\ \hline
{\rm Matter-dominated era} & $a(t) \sim t^{2/3}$ &
$a(t) \sim t^{\alpha}$, $(\alpha > 2/3)$ \\ \hline
{\rm Accelerating phase} & $a(t) = a(0)e^{\tilde{\chi}t}$ &
$a(t) = a(0)[1+\chi t]^{(2\omega +1)/2}$  \\ \hline
\end{tabular}
\caption{Effect of the presence of the BD scalar field on the universe's late-time evolution.} \label{t2}
\end{table}
Thus to conclude, the BD scalar field, which is a k-essence field of our model, appears to interpolate
{\it smoothly} between the decelerating matter-dominated era and the accelerating phase by reducing the
deceleration of the former somewhat and then reducing the acceleration of the latter to some extent.
And this role of ``crossing bridge'' between the two late-time epochs is indeed consistent with the result of 
our earlier study of BD scalar field as an unique k-essence in section II 
in which we idealized the situation by ignoring 
all other types of matter, namely, the emergence of a zero acceleration epoch in between the two late-time
eras. Thus we would like to point out that this smooth interpolation between the matter-dominated and the
accelerating eras is the key role played by the BD scalar field as an unique k-essence.  \\
(III) It is interesting to note that the BD parameter $\omega $ should ``run'' with the scale 
in order to serve as a successful model for dark matter - dark energy. Namely, according to the analysis
performed in subsections 1 and 2 of section III above, it appears that the BD $\omega$-parameter should
behave as a {\it running} coupling parameter with {\it growing} behavior
\begin{eqnarray}
\left\{%
\begin{array}{ll}
 \omega <-3/2 ~({\rm say}) & \hbox{{\rm in MATTER-TO-SCALAR transition period}, } \\
 \omega = -3/2 & \hbox{{\rm in (BD) scalar field-dominated era},} \\
 \omega > 1/2 & \hbox{{\rm in SCALAR-TO-ACCELERATION transition period}} \\
\end{array}%
\right.
\nonumber
\end{eqnarray}
as the scale factor $a(t)$ increases (with time). As we mentioned earlier, the BD parameter $\omega $ in
our model is a parameter of a k-essence (i.e., the BD scalar field) theory. Thus from a scalar field
theory perspective, this ``running'' behavior of the BD $\omega $-parameter can be viewed as being
natural in the sense of the renormalization group approach. What seems to be noteworthy is the fact that 
its value should grow with growing scale factor. We should, however, mention that this running behavior is
not a prediction of our model. An explicit demonstration of this running behavior,
although demanded, not only is beyond the scope of the present work but also seems unlikely to work out at the
level of microscopic field theory as the present study involves the macroscopic equations of state
in each epoch describing collective matter contents.  \\
(IV) Lastly, one may wonder if the presence of an additional epoch, i.e., the (BD) scalar field-dominated
era could be inconsistent with the current estimate (and observation by WMAP \cite{wmap}) of the age of
the universe based on the assumption that the universe is presently matter-dominated and should have been 
so for most of its history. Under this assumption, using $H_{0} = \dot{a}/a = \alpha /t$, 
$(\alpha > 2/3)$ for the matter-domination and its observed value $H_{0} \simeq 70 km ~s^{-1} ~Mpc^{-1}$
with $1 Mpc = 3.1\times 10^{24} cm$, the age of the universe is
estimated to be $t_{0} \simeq 10 ~(Gyr)$. As is now well-known from, say, the WMAP data \cite{wmap},
the universe energy density consists of contributions of the same order of magnitude from dark matter  
and dark energy. And this indicates that the present stage of
universe evolution is a sort of mixture of matter-dominated and accelerating eras. Furthermore, what we
have realized in this work is that there could be an additional intermediate epoch of zero acceleration 
in between these two eras which is the (BD) scalar field-dominated epoch. 
Therefore, the conventional picture of universe history above now should be modified accordingly and the 
new ingredient realized in the present work, i.e., the (BD) scalar field-dominated era could also be 
inserted in the late time. And it means durations of all the three (allowing, of course, for their 
possible overlaps) would sum up to the age of the universe. To date, the cosmic age based on the
best fit to a combination of WMAP, 2dFGRS (Two-Degree Field Galaxy Redshift Survey), Ly$\alpha $ forest data,
and the running index model is known to be $t_{0} \simeq 13.7 \pm 0.2 ~(Gyr)$ \cite{wmap}. 
In what follows, we attempt to work out this modification and to this end, we shall follow the formal
way of estimating the age of the universe given by Peebles and Ratra \cite{peebles2}.
Note first that even in the context of the BD theory, the Friedmann equation (in terms of the
spatially-flat FRW metric) can be put in the form
\begin{eqnarray}
\left(\frac{\dot{a}}{a}\right)^2 &=& \frac{8\pi G_{0}}{3}(\rho_{rad} + \rho_{m} + \rho_{BD} + \rho_{\Lambda}) 
= H^2_{0}E^2(z) ~~~{\rm where} \nonumber \\
E(z) &=& \left[\Omega_{rad0}(1+z)^4 + \Omega_{m0}(1+z)^3 + \Omega_{BD0}(1+z)^2 + \Omega_{\Lambda 0}\right]^{1/2},  \\
\Omega_{i0} &\equiv& \frac{8\pi G_{0}}{3H^2_{0}}\rho_{i0}, ~~~{\rm and} ~~~(1+z) = \frac{a(t_{0})}{a(t)}
\nonumber
\end{eqnarray}
with the subscript ``$0$'' indicating the present value of each quantity and $z$ the redshift.
That the Friedmann equation in BD theory can still be written as the first line in above equation (52)
results from the fact that the terms 
$\omega \left(\dot{\Phi}/\Phi\right)/6 - \left(\dot{a}/a\right)\left(\dot{\Phi}/\Phi\right)$, say,
in eqs.(36) and (44) can actually be rewritten in terms of $\left(\dot{a}/a\right)^2$ using the 
solutions for the BD scalar field $\Phi(t)$ and the scale factor $a(t)$. This then allows us to
write the Friedmann equation in BD theory as in the first line in (52). 
Next, the age of the universe can be given by
\begin{eqnarray}
t_{0} = \int^{t_{0}}_{0}\frac{da}{\dot{a}} =
H^{-1}_{0}\int^{\infty}_{z=0}\frac{dz}{(1+z)E(z)}.
\end{eqnarray}
Although the formal evaluation of this integral is intractable, one may still wish to get a rough
estimate of the cosmic age via the approximation
\begin{eqnarray}
t_{0} &\simeq& H^{-1}_{0}\left\{\int^{\infty}_{z^f_{rad}}\frac{dz}{(1+z)\left[\Omega_{rad0}(1+z)^4\right]^{1/2}}
+ \int^{z^{i}_{m}}_{z^f_{m}}\frac{dz}{(1+z)\left[\Omega_{m0}(1+z)^3\right]^{1/2}} \right. \nonumber \\
&+&\left. \int^{z^{i}_{BD}}_{z^f_{BD}}\frac{dz}{(1+z)\left[\Omega_{BD0}(1+z)^2\right]^{1/2}}
+ \int^{z^{i}_{\Lambda}}_{0}\frac{dz}{(1+z)\left[\Omega_{\Lambda0}\right]^{1/2}}\right\}  \\
&=& H^{-1}_{0}\left\{\frac{1}{2}\Omega^{-1/2}_{rad0}\frac{1}{(1+z^{f}_{rad})^2} +
\frac{2}{3}\Omega^{-1/2}_{m0}\left[\frac{1}{(1+z^{f}_{m})^{3/2}} - \frac{1}{(1+z^{i}_{m})^{3/2}}\right] \right. \nonumber \\
&+&\left. \Omega^{-1/2}_{BD0}\left[\frac{1}{(1+z^{f}_{BD})} - \frac{1}{(1+z^{i}_{BD})}\right] +
\Omega^{-1/2}_{\Lambda0}\ln (1+z^{i}_{\Lambda})\right\}.  \nonumber
\end{eqnarray}
The cosmological parameters in the standard LCDM model appearing in this expression for the cosmic age
have been determined from the observations mainly by WMAP \cite{wmap, peebles2} and they are
$\Omega_{rad0} \simeq 10^{-4}$, $\Omega_{m0}=\Omega_{vis0}+\Omega_{DM0} \simeq 0.05+0.20 = 0.25$,
$\Omega_{\Lambda0} \simeq 0.70$ for the fractional present energy densities and
$(z^{i}_{rad}\sim 10^{10}, ~z^{f}_{rad}\sim 10^3)$, $(z^{i}_{m}\sim 10^3, ~z^{f}_{m}\sim 1)$, and
$z^{i}_{\Lambda}\sim 0.5$ (particularly note that the late-time cosmic acceleration had begun at the
redshift of $z \sim 0.5$ or $6 ~Gyr$ ago). The actual evaluation of the cosmic age in the context of our
Brans-Dicke LCDM model based on eq.(54) above, however, is obscured as the data associated with the
BD scalar field-dominated era such as $\Omega_{BD0}$ and $(z^{i}_{BD}, ~z^{f}_{BD})$ are unknown yet.
In the most naive sense, however, we might wish to proceed with the rough estimate by assuming that
since $\Omega_{rad0}+\Omega_{m0}+\Omega_{BD0}+\Omega_{\Lambda0}=1$, from the LCDM data above, perhaps
$\Omega_{BD0} \leq 0.05$ and $(z^{i}_{BD}\sim 1, ~z^{f}_{BD}\sim 0.5)$ as the BD scalar field-dominated era
should be inserted in between the matter-dominated era and the current accelerating phase.
Lastly, then, by plugging these data numbers in eq.(54) above, we end up with an estimate for the age of
the universe in our Brans-Dicke LCDM model to be $t_{0} \simeq 1.63\times H^{-1}_{0} \simeq 23 ~(Gyr)$.
This is somewhat larger than the present estimate based on the best fit to observations. We hope, however,
that more sensible choice of the data, $\Omega_{BD0}$ and $(z^{i}_{BD}, ~z^{f}_{BD})$, if available at all,
would bring it down to some extent.

\begin{center}
{\rm\bf IV. Summary and discussion}
\end{center}

We now summarize our suggestion in the present work for adopting the BD theory of gravity as a 
``unified model'' for dark matter - dark energy. First, it seems worthy of note that it has never been 
shown at the level of fundamental physics that the dark matter and the dark energy should be two distinct 
ingredients. Indeed in most of the recent studies on the unified dark matter - dark energy, models employing 
a single component (whether it is the quintessence/k-essence or an exotic type of cosmic fluid) have been 
strongly favored. And needless to say, this is due to their attractive features of being able to provide 
an answer to the cosmic coincidence problem we mentioned earlier. In this regard, the way the BD theory
serves as a unified dark matter - dark energy model is indeed quite different from that other models
appeared in the recent literature do and thus is unique. Namely in our suggestion, first the BD scalar field
can be thought of as being qualified to play a role of a k-essence since it possesses perhaps the simplest
form of non-canonical kinetic term plus the interesting non-minimal coupling term to gravity.
In section II, it has been demonstrated that the BD scalar field-dominated era is a {\it zero acceleration}
epoch acting as a crossing bridge between the decelerating matter-dominated era and the accelerating phase.
This result is remarkable enough but has been based upon the idealization/simplification that in the
(BD) scalar field-dominated era, contributions from other types of matter may be dropped as they can be 
safely neglected. Therefore, in order to support and confirm this unique feature of the BD scalar field
as a k-essence, more careful and realistic treatment is called for and this has been actually attempted 
in section III. That is, since the nature of the late-time universe evolution appears to be some
mixture of matter-dominated, (BD) scalar field-dominated and accelerating phases, there we carried out
a closer study of the effects of the BD scalar field on the evolutionary behavior of the matter-dominated
and accelerating eras. And the result of the study in section III shows that the BD scalar field appears 
to interpolate {\it smoothly} between these two late-time stages by 
speeding up the expansion rate of the matter-dominated era somewhat while slowing down that of the 
accelerating phase to some degree. Thus with the newly found BD scalar field-dominated era in between 
these two, the late-time of the universe evolution appears to be mixed (allowing for possible overlaps)
sequence of the three stages.
Then our suggestion for employing the BD theory as a unified model for dark matter and dark energy can
be stated as follows. In the MATTER-TO-SCALAR transition period, the dust-like
matter and the BD scalar field together can be identified with some mixture of ordinary and dark matter and
in the SCALAR-TO-ACCELERATION transition period, the BD scalar field and the cosmological constant
$\Lambda$ together may be thought of as constituting some exotic type of dark energy. 
In other words, unlike the other ambitious models in the recent literature \cite{chaplygin, scherrer, microphysics}
attempting to build a unified dark matter - dark energy
in terms of a single component, the philosophy behind our model is to employ the BD scalar field as a
catalyzer that essentially generates the {\it dark} nature from the ordinary matter and the cosmological 
constant.  \\
We now turn to the comparison of the feature our BD scalar field-dominated universe studied in section II 
with that of the curvature-dominated universe in the FRW universe model of Einstein gravity. 
As is well-known, the curvature-dominated universe is (effectively)
an ``empty space'' solution in the spatially-open ($k=-1$) FRW model of Einstein gravity. As such,
the universe expansion occurs essentially due to the {\it negative} curvature of the spatial 
section ($k=-1$). That is, all types of matter are absent (or assumed to be negligible) and the
curvature term in the Friedmann equation behaves as if it is some kind of an energy density term,
$(\dot{a}/a)^2 = -k/a^2 = 1/a^2$ which gives $a(t)\sim t$. And the acceleration is zero simply
because $P=0=\rho $ in eq.(21), i.e., $(\ddot{a}/a) = -(4\pi G_{0}/3)(3P+\rho ) = 0$. \\
By contrast, the BD scalar field-dominated universe can be thought of as the k-essence-dominated
solution in the spatially-flat ($k=0$) FRW model of a non-minimally coupled Einstein-scalar theory
(or the pure BD theory of gravity). And the universe expansion occurs due to the k-essence
(i.e., the BD scalar field) energy density $\rho_{BD}\sim 1/a^2$ in the Friedmann equation,
$(\dot{a}/a)^2 = (8\pi G_{0}/3)\rho_{BD} = \tilde{\beta}^2a^2(0)/a^2$ (where we used eqs.(5),(11) and
(15)). Particularly the zero acceleration can be attributed to the {\it negative} pressure
$P_{BD}=-\rho_{BD}/3$ in eq.(24), i.e., $(\ddot{a}/a) = -(8\pi G_{0}/9)(3P_{BD}+\rho_{BD}) = 0$. 
To summarize, although the two universe models appear to exhibit the same expansion behavior
$a(t)\sim t$, $\ddot{a} = 0$, the origin/nature of zero acceleration of the former comes from the
{\it negative} spatial curvature (a geometrical nature) while that of the latter comes from the
{\it negative} pressure of the BD scalar field or k-essence (a matter).  Of course the essential
difference between the two arises from the fact that the spatial section of the universe is open for
the former model whereas it is flat for the latter model.    

\vspace*{1cm}

\begin{center}
{\rm\bf Acknowledgements}
\end{center}
The author would like to thank the anonymous referee for relevant criticisms and kind advices.

\end{document}